\documentstyle[amssymb,12pt,a4]{article}
\begin{document}
%\title{Dynamics of market indices, Markov chains, \\
%and random walking problem}
%\author{M. I. Krivoruchenko\thanks{%
%Permanent address: Institute for Theoretical and Experimental Physics, B.
%Cheremushkinskaya 25, 117259 Moscow, Russia}}
%\address{Institut f\"{u}r Theoretische Physik, Universit\"{a}t T\"{u}bingen\\
%Auf der Morgenstelle 14, D-72076 T\"{u}bingen, Germany}
%\maketitle
%{\small Institut f\"{u}r Theoretische Physik, Universit\"{a}t
%T\"{u}bingen, Auf der Morgenstelle 14}\\
%{\small D-72076 T\"{u}bingen, Germany}\\
%{\small Institute for Theoretical and Experimental Physics, B. Cheremushkinskaya 25}\\
%{\small 117259 Moscow, Russia}}
%\date{}
%\maketitle

%%%%%%
\title{Dynamics of market indices, Markov chains, \\
and random walking problem}
\author{M. I. Krivoruchenko \vspace{0.5 cm}{} \\
%EndAName
{\small \ }{\small Institute for Theoretical and Experimental
Physics, B. Cheremushkinskaya 25}\\
{\small 117259 Moscow, Russia}\\
{\small Institut f\"{u}r Theoretische Physik, Universit\"{a}t
T\"{u}bingen, Auf der Morgenstelle 14}\\
{\small D-72076 T\"{u}bingen, Germany}}
\date{}
\maketitle

%%%%%%%%%%%

\begin{abstract}
Dynamics of the major USA market indices DJIA, S$\&$P, Nasdaq, and NYSE is
analyzed from the point of view of the random walking problem with two-step
correlations of the market moves. The parameters characterizing the
stochastic dynamics are determined empirically from the historical quotes for
the daily, weekly, and monthly series. The results show existence of
statistically significant correlations between the subsequent market moves.
The weekly and monthly parameters are calculated in terms of the daily
parameters, assuming that the Markov chains with two-step correlations give
a complete description of the market stochastic dynamics. We show that the
macro- and micro-parameters obey the renorm group equation. 
The comparison of the parameters determined from the
renorm group equation with the historical values shows that the Markov
chains approach gives reasonable predictions for the weekly quotes and 
underestimates the probability for continuation of the down trend in 
the monthly quotes. The return and its dispersion for the
''buy-and-hold'' and ''follow-trend'' strategies are calculated. The problem
of how to combine these two strategies to reduce the dispersion is discussed
and its analytical solution is proposed. The results of the
constructing a "computer-based" strategy which combines the series analysis
with the candlesticks charting techniques are reported.
\end{abstract}
\newpage

\section{Introduction}
\setcounter{equation}{0}
$\;$
\vspace{-0.5cm}

At present, the economic data are easily accessible through Internet. The economic 
data are
renewed on the monthly, weekly, and daily basis. During the sessions the data are
renewed in the interactive regime, i.e. instantaneously. It is not a problem
to get an access to the economic data anymore. The problem is to make sense of the
available information.

During the last years, the possibilities for analyzing the data have been
improved very much. Now, the memory of personal computers is sufficient to
keep huge databases. The speed of the personal computers is
high enough to work with these databases and run complicated codes. Also, the
connection with Internet is fast enough to get within a reasonable time interval any
desirable information. This situation is completely new one, as compared
with the very recent past. 

The progress in the personal computers and in the Internet services improved
the possibilities of single programmers. One can expect that some of them
will attempt to understand the economic data, too.

The description of stochastic process requires the knowledge of the
probability theory and mathematical statistics. These mathematical tools are
useful not only in the exact sciences \cite{Chandrasekhar}, but also in the
social sciences \cite{Wilks,Portenko}.

This work represents an attempt to apply methods of the probability theory and
mathematical statistics for the analysis of the market dynamics. We are
interested in finding statistically significant deviations from the purely
stochastic market behavior. In particular, we are looking in historical quotes
for signatures which allow to predict market moves with sufficiently high 
probabilities. We use these signatures further to develop a computer-based 
strategy of the market behavior.

We start in the next Sect. from description of the random walking problem in
presence of a trend. This is an evident extension of the symmetric random
walking discussed by Chandrasekhar \cite{Chandrasekhar}. We calculate return
and dispersion for the ''buy-and-hold'' (BH)\ strategy in terms of a
trend parameter $w$ which represents a probability of moving an index up
for a day (week, month). Then, we determine from the historical quotes
empirical values of the trend parameters for the daily, weekly, and monthly
quotes of the four basic market indices of USA: Dow Jones Industrial Average
(DJIA, analyzed years 1930 - 2000), Standard and Poor (S$\&$P, analyzed
years 1950 - 2000), Nasdaq (analyzed years 1985 - 2000), and New York Stock
Exchange (NYSE, analyzed years 1965 - 2000).

In Sect. 3, we propose a more detailed model for dynamics of the market
indices by taking into account two-step correlations of the subsequent
market moves. Such a model represents a particular case of the Markov chains
that have many useful applications in different fields. The model has two
free correlation parameters, $p$ and $q$, which are probabilities for a {\it %
continuation} of the market moves, respectively, up and down. We determine
empirical values of these parameters from the historical quotes for the daily,
weekly, and monthly data of the four basic market indices. These two-step
correlations are found to be statistically significant. We calculate further
return and dispersion for the BH and ''follow-trend'' (FT) strategies
and compare their efficiency in the past. The
problem of how to combine these two strategies to ensure the minimum of the
dispersion (a minimum risk strategy) has an analytical solution.
We found it using the theory of series \cite{Wilks}. 

In Sect. 4, we calculate within the framework of the Markov chains approach
the weekly and monthly correlation parameters $P$ and $Q$ in terms of the
daily correlation parameters $p$ and $q$. The results are compared with the
empirical values. There are deviations of the calculated values from the
empirical values, which indicate that the large-scale dynamics of the market
indices is not fully determined by the short-scale dynamics. The
renorm group relations are established between the correlation parameters at
different time scales.

In Sect. 5, backtesting principles are discussed.
We show that misuse of the backtesting can lead to systematic overestimates of the
returns and give an example of such a misuse. We propose a special type of the
backtesting, which is more realistic with respect to the
estimate of the future performance. In this Sect., 
specific features of the market dynamics are used to
construct a computer-based (CB) strategy based on a combination of the theory
of series and Japanese candlesticks charting techniques. The CB strategy is
backtested. 

The results obtained in this work are discussed in Conclusion.

\section{Random walking with trend}
\setcounter{equation}{0}
$\;$
\vspace{-0.5cm}

An excellent introduction to the random walking problem can be found in the
Chandrasekhar's lectures \cite{Chandrasekhar}.

We consider random walking of a particle on a vertical infinite line. On
each step, the particle makes a move up with a probability $w$ or a move
down with a probability $1-w$. The length of the jumps is everywhere the same
and set equal to unity.

In order to arrive after $N$ steps into a point with a coordinate 
\begin{equation}
z=n_{+}-n_{-},  \label{z}
\end{equation}
the particle should make $n_{+}=N-n_{-}$ steps up and $n_{-}$ steps down.
The probability of finding the particle with a coordinate $z$ after $N$
steps is described by the binomial distribution 
\begin{equation}
W(n_{+},n_{-})=\frac{N!}{n_{+}!n_{-}!}w^{n_{+}}(1-w)^{n_{-}}.  \label{binom}
\end{equation}
This distribution is normalized to unity: 
\begin{equation}
\sum_{n_{-}=0}^NW(N-n_{-},n_{-})=1.  \label{norm}
\end{equation}

The dynamics of the market indices can be treated from the point of view of
the random walking problem. If a market index is closed up for a day
(week, month, ...), we say that the ''particle'' made a step up. If a market
index is closed down, we say that the ''particle'' made a step down.

We analyzed the following data: DJIA (1930 - 2000 years), S$\&$P (1950 -
2000), Nasdaq (1985 - 2000), and NYSE (1965 - 2000). In Table 1, the values
of $w$ are shown for the daily ($W_1$), weekly ($W_5$), and monthly ($W_{21}$%
) quotes of these indices. The meaning of other parameters in Table 1 is
explained in the next Sects. In the meantime notice that the values of $w$
everywhere are quite close to $1/2$. This case of the random walking is discussed
by Chandrasekhar \cite{Chandrasekhar}.

As compared to his lectures, we discuss more general case and introduce a
trend parameter $w$ that can be different from $1/2$. The deviation of the
daily parameter $w$ from the symmetric value $1/2$ is statistically
significant. Indeed, the accuracy of the numbers given in Table 1 is
determined for the daily quotes by the value of $1/\sqrt{N_D}$ where $N_D=($the
number of the working days per year$)\times ($the number of years analyzed$%
). $ For DJIA index, the statistical error equals $0.7\%$. The deviation of
the daily parameter $w=0.5339$ from the symmetric value $w=1/2$ is $6.78\%$
i.e. $10$ times greater. It is safe to conclude therefore that the trend
does exist. The statistical errors are shown in Table 1 for all indices for
the daily, weekly, and monthly quotes. Notice that according to the estimate
of ref. \cite{Colby}, p. 141, based on the analysis of the 1952 - 1983
years, the DJIA daily trend parameter equals $w=0.521$. This result is close
to ours.

{\tiny
\begin{table}
\caption{Trend parameters $W_k$ and correlation parameters $P_k$ and $%
Q_k$ for the daily, $k=1$, weekly, $k=5$, and monthly, $k=21$, data for the
four major USA market indices. The empirical values of the parameters are
determined by fitting the distributions of the number of the series with the
growing and decreasing moves assuming the homogeneous condition (3.2) is
fulfilled for the daily, weekly, and monthly data. The statistical errors of
the parameters are shown. The calculated values of the parameters are found
assuming that the large-scale dynamics is determined by the short-scale
dynamics as described in Sect. 5. For $k=1,$ the parameters $w=W_1$, $p=P_1$%
, and $q=Q_1$ are used as an input.}
\begin{center}
\begin{tabular}{|ccccccccc|}
\hline\hline
\multicolumn{1}{|c|}{Parameters} & \multicolumn{2}{|c}{DJIA} & 
\multicolumn{2}{|c}{S$\&$P} & \multicolumn{2}{|c}{Nasdaq} & 
\multicolumn{2}{|c|}{NYSE} \\ 
\cline{3-3}\cline{2-2}\cline{5-5}\cline{4-4}\cline{6-6}\cline{7-7}\cline{8-8}\cline{9-9}
\multicolumn{1}{|c|}{} & \multicolumn{1}{c|}{calculated} & 
\multicolumn{1}{c|}{empirical} & \multicolumn{1}{c|}{calculated} & 
\multicolumn{1}{c|}{empirical} & \multicolumn{1}{c|}{calculated} & 
\multicolumn{1}{c|}{empirical} & \multicolumn{1}{c|}{calculated} & empirical
\\ \hline\hline
\multicolumn{1}{|c|}{$W_1$} & \multicolumn{1}{c|}{\bf input} & 
\multicolumn{1}{c|}{$0.5339\pm 0.0037$} & \multicolumn{1}{c|}{\bf input} & 
\multicolumn{1}{c|}{$0.5382\pm 0.0044$} & \multicolumn{1}{c|}{\bf input} & 
\multicolumn{1}{c|}{$0.5829\pm 0.0080$} & \multicolumn{1}{c|}{\bf input} & $%
0.5302\pm 0.0054$ \\ 
\multicolumn{1}{|c|}{$P_1$} & \multicolumn{1}{c|}{\bf input} & 
\multicolumn{1}{c|}{$0.5698\pm 0.0045$} & \multicolumn{1}{c|}{\bf input} & 
\multicolumn{1}{c|}{$0.5918\pm 0.0059$} & \multicolumn{1}{c|}{\bf input} & 
\multicolumn{1}{c|}{$0.6579\pm 0.0129$} & \multicolumn{1}{c|}{\bf input} & $%
0.5886\pm 0.0072$ \\ 
\multicolumn{1}{|c|}{$Q_1$} & \multicolumn{1}{c|}{\bf input} & 
\multicolumn{1}{c|}{$0.5073\pm 0.0040$} & \multicolumn{1}{c|}{\bf input} & 
\multicolumn{1}{c|}{$0.5244\pm 0.0052$} & \multicolumn{1}{c|}{\bf input} & 
\multicolumn{1}{c|}{$0.5220\pm 0.0102$} & \multicolumn{1}{c|}{\bf input} & $%
0.5357\pm 0.0066$ \\ \hline\hline
\multicolumn{1}{|c|}{$W_5$} & \multicolumn{1}{c|}{$0.5596$} & 
\multicolumn{1}{c|}{$0.5623\pm 0.0082$} & \multicolumn{1}{c|}{$0.5650$} & 
\multicolumn{1}{c|}{$0.5726\pm 0.0097$} & \multicolumn{1}{c|}{$0.6331$} & 
\multicolumn{1}{c|}{$0.6018\pm 0.0175$} & \multicolumn{1}{c|}{$0.5513$} & $%
0.5579\pm 0.0118$ \\ 
\multicolumn{1}{|c|}{$P_5$} & \multicolumn{1}{c|}{$0.5644$} & 
\multicolumn{1}{c|}{$0.5784\pm 0.0099$} & \multicolumn{1}{c|}{$0.5722$} & 
\multicolumn{1}{c|}{$0.5893\pm 0.0120$} & \multicolumn{1}{c|}{$0.6425$} & 
\multicolumn{1}{c|}{$0.6417\pm 0.0255$} & \multicolumn{1}{c|}{$0.5593$} & $%
0.5800\pm 0.0145$ \\ 
\multicolumn{1}{|c|}{$Q_5$} & \multicolumn{1}{c|}{$0.4465$} & 
\multicolumn{1}{c|}{$0.4585\pm 0.0079$} & \multicolumn{1}{c|}{$0.4443$} & 
\multicolumn{1}{c|}{$0.4499\pm 0.0092$} & \multicolumn{1}{c|}{$0.3832$} & 
\multicolumn{1}{c|}{$0.4585\pm 0.0182$} & \multicolumn{1}{c|}{$0.4585$} & $%
0.4699\pm 0.0117$ \\ \hline\hline
\multicolumn{1}{|c|}{$W_{21}$} & \multicolumn{1}{c|}{$0.6150$} & 
\multicolumn{1}{c|}{$0.5815\pm 0.0170$} & \multicolumn{1}{c|}{$0.6244$} & 
\multicolumn{1}{c|}{$0.5929\pm 0.0200$} & \multicolumn{1}{c|}{$0.7423$} & 
\multicolumn{1}{c|}{$0.6589\pm 0.0353$} & \multicolumn{1}{c|}{$0.5984$} & $%
0.5821\pm 0.0244$ \\ 
\multicolumn{1}{|c|}{$P_{21}$} & \multicolumn{1}{c|}{$0.6159$} & 
\multicolumn{1}{c|}{$0.6012\pm 0.0218$} & \multicolumn{1}{c|}{$0.6257$} & 
\multicolumn{1}{c|}{$0.6379\pm 0.0293$} & \multicolumn{1}{c|}{$0.7436$} & 
\multicolumn{1}{c|}{$0.6604\pm 0.0497$} & \multicolumn{1}{c|}{$0.6000$} & $%
0.6288\pm 0.0350$ \\ 
\multicolumn{1}{|c|}{$Q_{21}$} & \multicolumn{1}{c|}{$0.3364$} & 
\multicolumn{1}{c|}{$0.4458\pm 0.0162$} & \multicolumn{1}{c|}{$0.3778$} & 
\multicolumn{1}{c|}{$0.4727\pm 0.0217$} & \multicolumn{1}{c|}{$0.2614$} & 
\multicolumn{1}{c|}{$0.3439\pm 0.0259$} & \multicolumn{1}{c|}{$0.4039$} & $%
0.4830\pm 0.0269$ \\ 
\hline\hline
\end{tabular}
\end{center}
\end{table}
}

We analyze finite samples of the data,
and so an attempt to find a statistical law can always result in finding a
statistical fluctuation. Such an unwanted possibility always exists and
cannot be excluded. Maximum what can be done is to reduce risk of the
misinterpreting the data. In addition, we deal with an alive system. Even if
it is described by a statistical law, parameters of the corresponding
distribution can be time dependent. If the rate of accumulation of the data
(day-by-day, week-by-week, ...), which we clearly cannot influence, is lower
than the rate of variation of the system parameters, the system parameters
cannot be determined from the observations. There always exists an uncertainty.

The statistical methods are, however, widely used in the social sciences, 
despite they are less reliable than in the exact sciences.
In this work, we want to reach some clarity concerning a few basic features
of the market behavior in the past. We do not discuss if these feature
will remain in the future markets.

The very important characteristics of a market index is its return. In our
simplified model, return is nothing but the number of the moves up minus the
number of the moves down. This quantity looks like 
\begin{equation}
R_1/N=2w-1.  \label{R1}
\end{equation}
We have here the difference of the steps up ($w$) and down ($1-w$). This is,
equivalently, the average value of the parameter $z=n_{-}-n_{+}$ in the
binomial distribution (\ref{binom}).

We refer this passive strategy as ''buy-and-hold'' strategy. Notice that 
\begin{eqnarray}
&<&n_{-}>=(1-w)N,  \label{2.4} \\
&<&n_{-}^2>=w(1-w)N+(1-w)^2N^2,  \label{2.5}
\end{eqnarray}
and so the dispersion of the value $z=n_{-}-n_{+}$ equals 
\begin{equation}
\sigma _1^2=4w(1-w)N.  \label{S1}
\end{equation}

\section{Random walking with two-step correlations}
\setcounter{equation}{0}
$\;$
\vspace{-0.5cm}

In order to reveal less trivial statistical features of the market dynamics,
we should consider a possibility for existence of correlations between the 
subsequent moves. In other words, we want to check if 
there is a memory about past when an index makes a move.

In the simplest case, the memory effect is described by four conventional
probabilities 
\begin{eqnarray}
&&w(\uparrow |\uparrow )=p,  \nonumber \\
&&w(\downarrow |\uparrow )=1-p,  \nonumber \\
&&w(\uparrow |\downarrow )=1-q,  \nonumber \\
&&w(\downarrow |\downarrow )=q,  \label{p.and.q}
\end{eqnarray}
which depend on two independent parameters $p$ and $q$. We refer these
parameters as correlation parameters. The signs $\uparrow $ and $\downarrow $
denote moves up and down.

We would like to deal with a homogeneous system with respect to the time
translations. It means that any particular sample of the data should give
identical results for the trend parameter $w$ and the correlation parameters 
$p$ and $q$. The homogeneous condition looks like 
\begin{equation}
w=pw+(1-q)(1-w).  \label{homo}
\end{equation}
It shows that the value of $w$ remains invariant under the time
translations. It becomes evident if eq.(\ref{homo}) is rewritten in the form 
\[
w(\uparrow )=w(\uparrow |\uparrow )w(\uparrow )+w(\uparrow |\downarrow
)w(\downarrow ). 
\]
Here, $w(\uparrow )=w$ and $w(\downarrow )=1-w$, while the conventional
probabilities are from eqs.(\ref{p.and.q}).

Therefore, as long as the parameters $p$ and $q$ are known, the trend
parameter is fixed: 
\begin{equation}
w=\frac{1-q}{2-p-q}.  \label{3.2}
\end{equation}

If the homogeneous condition (\ref{homo}) is not fulfilled, the problem
makes sense, too. The difference is only that the parameters $p$ and $q$
determined from the first half of the data could be different from the same
parameters determined from the second half of the data. Such a possibility
is not discussed here.

In Table 1, the trend parameters $W_k$ and the correlation parameters $P_k$ and $%
Q_k$ are shown for $k=1$ (daily data), $5$ (weekly data), and $21$ (monthly
data) for the four major USA market indices. The empirical values are
determined by fitting the distribution of number of the sequential series
for the up- and down-moves under the imposed homogeneity condition (\ref
{homo}). The statistical errors are also shown. The meaning of the
''calculated values'' $W_k$, $P_k$, and $Q_k$ is explained in Sect. 4. The
weekly and monthly data are discussed in the details in Sect. 4.

The main observation that can be made at the moment is the existence of the
correlations between two consequent steps of the market indices in the daily
data. These correlations are not strong, but statistically significant. The lack
of the correlations would mean $p + q = 1$ and $w=p$. To
the first approximation, the market dynamics can be treated as the random
walking without trend ($w=1/2$). To the next approximation, it can be
treated as the random walking with a trend ($w\neq 1/2$). The next
approximation takes into account the two-step correlations, and these
correlations are also statistically significant.

\begin{table}
\caption{Comparison of the $\chi ^2$ for the models without trend ($%
w=1/2$, No. 1), with a trend ($w\ne 1/2$, No. 2), and with the correlation for
the up- and down-events (No. 3), respectively, for the daily (D), weekly (W), and
monthly (M) quotes of the four major USA market indices.}

\begin{center}
\begin{tabular}{|c|c|c|c|c|c|c|c|c|c|}
\hline\hline
Type & No. & \multicolumn{2}{|c}{DJIA} & \multicolumn{2}{|c}{S$\&$P} & 
\multicolumn{2}{|c}{Nasdaq} & \multicolumn{2}{|c|}{NYSE} \\ 
\cline{3-3}\cline{4-4}\cline{5-5}\cline{6-6}\cline{7-7}\cline{8-8}\cline{9-9}\cline{10-10}
&  & up & down & up & down & up & down & up & down \\ \hline\hline
& 1 & $196.7$ & $80.02$ & $192.2$ & $23.51$ & $137.9$ & $4.59$ & $118.7$ & $%
79.40$ \\ 
D & 2 & $78.87$ & $83.21$ & $79.49$ & $87.41$ & $42.04$ & $50.5$ & $59.53$ & 
$53.81$ \\ 
& 3 & $40.32$ & $96.47$ & $13.36$ & $14.90$ & $12.30$ & $3.07$ & $11.59$ & $%
20.60$ \\ \hline\hline
& 1 & $37.95$ & $27.9$ & $34.3$ & $18.2$ & $20.9$ & $5.83$ & $18.46$ & $11.9$
\\ 
W & 2 & $10.74$ & $8.96$ & $8.14$ & $2.77$ & $9.04$ & $3.45$ & $12.70$ & $%
2.56$ \\ 
& 3 & $12.64$ & $9.32$ & $8.69$ & $2.05$ & $9.78$ & $2.10$ & $16.77$ & $3.91$
\\ \hline\hline
& 1 & $11.61$ & $23.4$ & $26.26$ & $20.3$ & $6.95$ & $2.02$ & $13.47$ & $%
6.44 $ \\ 
M & 2 & $14.28$ & $5.95$ & $41.72$ & $7.13$ & $3.19$ & $0.84$ & $12.53$ & $%
3.66$ \\ 
& 3 & $18.48$ & $9.38$ & $59.46$ & $14.6$ & $3.20$ & $0.90$ & $25.02$ & $%
5.30 $ \\ 
\hline\hline
\end{tabular}
\end{center}
\end{table}

In Table 2, we compare $\chi ^2$ for the ratios between the empirical number
of the up- and down-series and the number of the analogous series for the random 
walking in three models: The symmetric random walking $w=1/2$, the random walking 
with trend ($w \ne 1/2$), and the random walking with trend and 
correlations. It is seen that for the daily quotes the $\chi ^2$ for the third 
model is 5 to 15 times less than for the first two models for the up-series and is
comparable with the first two models for the down-series. We consider it
as a clear evidence for the existence of the two-step correlations. 
The equation $p+q=1$ can also be treated as a criterion for the existence of the
two-step correlations. For the daily quotes, it is not fulfilled, 
as one sees from the Table 1. The similar check can also be made
for the equation $w=p$ which is an equivalent criterion for the existence of the
correlations.

There are deviations from the geometrical distribution of the number of the series 
of a fixed length. 
In Tables 3, 4, and 5, we show the daily (D), weekly (W), and monthly (M) 
empirical values $p_k^{*}$ together with the statistical errors for 
probabilities of continuation of the series 
\begin{equation}
p_k^{*}=\frac{n_k+n_{k+1}+...}{n_{k-1}+n_k+...},  \label{p_star}
\end{equation}
with $n_k\;$being the number of the up-series of the length $k$. So, if we
have $k-1$ moves up, the probability to have the next move up is $p_k^{*}.$
The similar expression holds true for the values $q_k^{*}$ that describe the
probability for continuation of the series down. The results are shown for
the four basic market indices of USA. The numbers typed in boldface deviate
more than others from the symmetric values $p = q = 1/2$. They approach 1,
signaling continuation of the series with a high probability, or approach 0, 
signaling the end of the series with a high probability. 

{\tiny
\begin{table}
\caption{Empirical values of $p_k^{*}\;$and $q_k^{*}\ $for daily
quotes.}
\begin{center}
\begin{tabular}{||c||c||cccccc||}
\hline\hline
${\rm Index}$ & $k$ & $2$ & $3$ & $4$ & $5$ & $6$ & $7$ \\ \hline\hline
DJIA & $p^{*}$ & $0.5947\pm .0148$ & $0.5570\pm 0.0184$ & $0.5196\pm 0.0235$
& $0.5708\pm 0.0347$ & $0.5224\pm 0.0433$ & $0.5022\pm 0.0584$ \\ 
& $q^{*}$ & $0.5198\pm 0.0135$ & $0.4986\pm 0.0182$ & $0.4892\pm 0.0255$ & $%
0.4816\pm 0.0361$ & $0.4106\pm 0.0469$ & $0.4166\pm 0.0739$ \\ 
S$\&$P & $p^{*}$ & $0.6087\pm 0.0186$ & $0.5882\pm 0.0234$ & $0.5553\pm
0.0293$ & $0.5691\pm 0.0400$ & $0.5930\pm 0.0545$ & $0.5585\pm 0.0680$ \\ 
& $q^{*}$ & $0.5421\pm 0.0172$ & $0.5075\pm 0.0224$ & $0.4928\pm 0.0308$ & $%
0.5131\pm 0.0452$ & $0.5025\pm 0.0622$ & $0.4183\pm 0.0778$ \\ 
Nasdaq & $p^{*}$ & ${\bf 0.6532\pm 0.0373}$ & ${\bf 0.6831\pm 0.0477}$ & $%
0.6144\pm 0.0536$ & $0.6179\pm 0.0686$ & $0.6183\pm 0.0873$ & ${\bf %
0.7160\pm 0.1231}$ \\ 
& $q^{*}$ & $0.5271\pm 0.0322$ & $0.5122\pm 0.0435$ & $0.4928\pm 0.0593$ & $%
0.4563\pm 0.0803$ & $0.4680\pm 0.1209$ & $0.5454\pm 0.1957$ \\ 
NYSE & $p^{*}$ & $0.5984\pm 0.0225$ & $0.6028\pm 0.0292$ & $0.5485\pm 0.0353$
& $0.5683\pm 0.0488$ & $0.5566\pm 0.0639$ & $0.5254\pm 0.0824$ \\ 
& $q^{*}$ & $0.5323\pm 0.0207$ & $0.5418\pm 0.0288$ & $0.5238\pm 0.0383$ & $%
0.5298\pm 0.0533$ & $0.5364\pm 0.0738$ & $0.4320\pm 0.0874$ \\ \hline\hline
\end{tabular}
\end{center}
\end{table}
}

{\tiny
\begin{table}
\caption{Empirical values of $p_k^{*}\;$and $q_k^{*}\ $for
weekly quotes.}
\begin{center}
\begin{tabular}{||c||c||cccccc||}
\hline\hline
${\rm Index}$ & $k$ & $2$ & $3$ & $4$ & $5$ & $6$ & $7$ \\ \hline\hline
DJIA & $p^{*}$ & $0.5688\pm 0.0318$ & $0.5920\pm 0.0434$ & $0.5709\pm 0.0550$
& $0.5562\pm 0.0715$ & $0.4574\pm 0.0842$ & $0.6046\pm 0.1502$ \\ 
& $q^{*}$ & $0.4323\pm 0.0265$ & $0.4973\pm 0.0442$ & $0.4391\pm 0.0578$ & $%
0.4939\pm 0.0942$ & $0.4634\pm 0.1286$ & ${\bf 0.2105\pm 0.1158}$ \\ 
S$\&$P & $p^{*}$ & $0.5852\pm 0.0386$ & $0.5631\pm 0.0491$ & $0.6097\pm
0.0691$ & $0.6320\pm 0.0908$ & $0.4936\pm 0.0966$ & $0.6153\pm 0.1596$ \\ 
& $q^{*}$ & $0.4356\pm 0.0317$ & $0.4575\pm 0.0496$ & $0.4758\pm 0.0752$ & $%
0.3898\pm 0.0958$ & $0.4347\pm 0.1646$ & ${\bf 0.3000\pm 0.1974}$ \\ 
Nasdaq & $p^{*}$ & $0.5872\pm 0.0736$ & ${\bf 0.6732\pm 0.1056}$ & $%
0.5735\pm 0.1152$ & ${\bf 0.7435\pm 0.1823}$ & ${\bf 0.6896\pm 0.2004}$ & $%
{\bf 0.7000\pm 0.2439}$ \\ 
& $q^{*}$ & $0.4277\pm 0.0594$ & $0.4864\pm 0.0988$ & $0.4444\pm 0.1335$ & $%
{\bf 0.3125\pm 0.1601}$ & ${\bf 0.2000\pm 0.2190}$ &  \\ 
NYSE & $p^{*}$ & $0.5454\pm 0.0443$ & $0.5769\pm 0.0623$ & $0.6148\pm 0.0857$
& $0.6144\pm 0.1093$ & ${\bf 0.3725\pm 0.1001}$ & $0.6842\pm 0.2462$ \\ 
& $q^{*}$ & $0.4465\pm 0.0387$ & $0.4635\pm 0.0594$ & $0.4943\pm 0.0911$ & $%
0.3863\pm 0.1103$ & $0.4117\pm 0.1849$ & ${\bf 0.2857\pm 0.2290}$ \\ \hline\hline
\end{tabular}
\end{center}
\end{table}
}

{\tiny
\begin{table}
\caption{Empirical values of $p_k^{*}\;$and $q_k^{*}\ $for
monthly quotes.}
\begin{center}
\begin{tabular}{||cccccccc||}
\hline\hline
\multicolumn{1}{||c||}{${\rm Index}$} & \multicolumn{1}{c||}{$k$} & $2$ & $3$
& $4$ & $5$ & $6$ & $7$ \\ \hline\hline
\multicolumn{1}{||c||}{DJIA} & \multicolumn{1}{c||}{$p^{*}$} & $0.5741\pm
0.0657$ & $0.5333\pm 0.0825$ & $0.4843\pm 0.1059$ & $0.5806\pm 0.1720$ & $%
{\bf 0.7777\pm 0.2771}$ & $0.6428\pm 0.2746$ \\ 
\multicolumn{1}{||c||}{} & \multicolumn{1}{c||}{$q^{*}$} & $0.3666\pm 0.0488$
& $0.4545\pm 0.0926$ & $0.3714\pm 0.1206$ & $0.5384\pm 0.2524$ & $0.5714\pm
0.3581$ & ${\bf 0.2500\pm 0.2795}$ \\ 
\multicolumn{1}{||c||}{S$\&$P} & \multicolumn{1}{c||}{$p^{*}$} & $0.6187\pm
0.0848$ & $0.5116\pm 0.0948$ & $0.6136\pm 0.1500$ & $0.5925\pm 0.1869$ & $%
{\bf 0.9375\pm 0.3369}$ & $0.6666\pm 0.2721$ \\ 
\multicolumn{1}{||c||}{} & \multicolumn{1}{c||}{$q^{*}$} & $0.3785\pm 0.0610$
& $0.4905\pm 0.1174$ & $0.3461\pm 0.1338$ & $0.6666\pm 0.3513$ & $0.6666\pm
0.4303$ & ${\bf 0.2500\pm 0.2795}$ \\ 
\multicolumn{1}{||c||}{Nasdaq} & \multicolumn{1}{c||}{$p^{*}$} & $0.5853\pm
0.1504$ & ${\bf 0.7083\pm 0.2245}$ & ${\bf 0.7058\pm 0.2661}$ & ${\bf %
0.6666\pm 0.3042}$ & $0.5000\pm 0.3061$ & $0.7500\pm 0.5728$ \\ 
\multicolumn{1}{||c||}{} & \multicolumn{1}{c||}{$q^{*}$} & ${\bf 0.3095\pm
0.0982}$ & ${\bf 0.2307\pm 0.1478}$ & $0.3333\pm 0.3849$ &  &  & \\ 
\multicolumn{1}{||c||}{NYSE} & \multicolumn{1}{c||}{$p^{*}$} & $0.6063\pm
0.1017$ & $0.5087\pm 0.1160$ & $0.5862\pm 0.1790$ & ${\bf 0.7058\pm 0.2661}$
& ${\bf 0.9166\pm 0.3826}$ & $0.6363\pm 0.3076$ \\ 
\multicolumn{1}{||c||}{} & \multicolumn{1}{c||}{$q^{*}$} & $0.3936\pm 0.0763$
& $0.5405\pm 0.1500$ & $0.4000\pm 0.1673$ & $0.5000\pm 0.3061$ & $0.5000\pm
0.4330$ & $0.5000\pm 0.6123$ \\ \hline\hline
\end{tabular}
\end{center}
\end{table}
}

Notice that Nasdaq during the 1985 - 2000 years did not fall off
more than 6 weeks in raw in the weekly quotes and more than 4 month in raw in
the monthly quotes. Last time Nasdaq was down 4 months in raw in the period
September 2000 - December 2000. Using the results of Table 5, one could
predict with certainty (with unit probability) a positive move of Nasdaq 
for January 2001. Such a prediction would be the correct one.

Let us calculate return of the follow-trend (FT) strategy: We buy an index the next day when it
is closed up for the first time after one or several days down and sell it the
next day when it is closed down for the first time after one or several
days up. The return equals 
\begin{equation}
R_2/N=(2p-1)w+(2q-1)(1-w).  \label{R2}
\end{equation}
In order to derive eq.(\ref{R2}), one should consider the problem in more
details:

In the sample of all events, there are ${\frak r}_{+j}$ series of the length 
$j$, in which all $j$ events are moves up.
Similarly, there exist ${\frak r}_{-j}$ series of the length $j$, in which
all $j$ events represent moves down. In the both cases, the value of $j$
varies from $1$ to $+\infty ,$ in the limit of $N\rightarrow +\infty $. In what
follows, we set $N=+\infty .$ The finite values of $N$ if appear imply the
leading order at $N\rightarrow +\infty $.

The probability of finding up-series of the length $j$ in the set of all
up-series is apparently described by the geometrical distribution 
\begin{equation}
{\frak w}_{+j}=(1-p)p^{j-1}.  \label{w.p}
\end{equation}
This distribution is normalized to unity. The same distribution is valid for
description of the down-series of the length $j$ in the set of all down-series 
\begin{equation}
{\frak w}_{-j}=(1-q)q^{j-1}.  \label{w.m}
\end{equation}
The average number of events in the series can easily be found to be 
\begin{eqnarray}
\ &<&j_{+}>=\sum_{j=1}^\infty j{\frak w}_{+j}=\frac 1{1-p},  \label{aj.p} \\
\ &<&j_{-}>=\sum_{j=1}^\infty j{\frak w}_{-j}=\frac 1{1-q}.  \label{aj.m}
\end{eqnarray}
It is evident that the number of series ${\frak r}_{\pm j}$ of the length $j$
in a sufficiently large sample is proportional to the probability of finding
the same series, so one can write 
\begin{equation}
{\frak r}_{\pm j}=C_{\pm }{\frak w}_{\pm j}.  \label{r.pm}
\end{equation}
Since the number of all up- and down-moves is known, 
\begin{eqnarray}
\ &<&n_{+}>=\sum_{j=1}^\infty j{\frak r}_{+j}=\frac{C_{+}}{1-p}=wN,
\label{an.p} \\
\ &<&n_{-}>=\sum_{j=1}^\infty j{\frak r}_{-j}=\frac{C_{+}}{1-q}=(1-w)N,
\label{an.m}
\end{eqnarray}
where $<n_{+}>+<n_{+}>=N,$ the coefficients $C_{\pm }$ appear to be
determined. The average number of the series of the first and second kind
equals 
\begin{eqnarray}
\ &<&{\frak r}_{+}>=\sum_{j=1}^\infty {\frak r}_{+j}=\frac{wN}{<j_{+}>}%
=(1-p)wN,  \label{ar.p} \\
\ &<&{\frak r}_{-}>=\sum_{j=1}^\infty {\frak r}_{-j}=\frac{(1-w)N}{<j_{-}>}%
=(1-q)(1-w)N.  \label{ar.m}
\end{eqnarray}
Notice that due to the homogeneity condition (\ref{homo}), 
\begin{equation}
<{\frak r}_{+j}>=<{\frak r}_{-j}>.  \label{3.11}
\end{equation}
The return for the FT strategy represents a sum of the number of the series $%
{\frak r}_{+j}$ multiplied by the return $j-2$ of the $j$-th series plus
the analogous term for the down-series: 
\begin{equation}
R_2=\sum_{j=1}^\infty (j-2){\frak r}_{+j}+\sum_{j=1}^\infty (j-2){\frak r}%
_{-j}=N-2(<{\frak r}_{+}>+<{\frak r}_{-}>).  \label{3.12}
\end{equation}
With the help of eqs.(\ref{ar.p}) and (\ref{ar.m}) we obtain eq.(\ref{R2}).

In order to find the dispersion of the value $R_2$, it is necessary to perform
additional constructions in the spirit of the theory of series encountered
in the mathematical statistics (see e.g. \cite{Wilks}). 
A part of the necessary constructions is already done above. The additional
efforts are justified since the dispersion represents an important
characteristics of every strategy.

The value of $N$ is assumed to be large and
finite. The probability of finding a sample characterized by the values ${\frak r}%
_{+j}$ and ${\frak r}_{-j}$ can be written as follows 
\begin{equation}
W({\frak r}_{+j},{\frak r}_{-j},N)=C_N\frac{{\frak r}_{+}!{\frak r}_{-}!}{%
{\frak r}_{+1}!{\frak r}_{-2}!\;...\;{\frak r}_{-1}!{\frak r}_{-2}!\;...}%
p^{n_{+}}q^{n_{-}}(\frac{1-q}p)^{{\frak r}_{+}}(\frac{1-p}q)^{{\frak r}_{-}}.
\label{W}
\end{equation}
The combinatorial factor is the number of different ways to order ${\frak r}%
_{+}$ up- and ${\frak r}_{-}$ down-series in the sample
characterized by the sets of the numbers ${\frak r}_{+j}$ and ${\frak r}%
_{-j}.$ The combinatorial factor is multiplied further to a probability of
finding the one concrete configuration.

Each sample contains $n_{+}$ up- and $n_{-}$ down-moves. The
value $p^{n_{+}}$ is, in principle, a probability of finding $n_{+}$ up-moves. 
However, the first moves of the series appear every time with
probability of $1-q.$ There are ${\frak r}_{+}$ such moves, one move
per one series. The correct probability to have $n_{+}$ up-moves is
therefore $p^{n_{+}\;-\;{\frak r}_{+}}(1-q)^{{\frak r}_{+}}$ and similarly
for the down-moves. In this way, the rest terms in eq.(\ref{W}) are
reproduced.

The very first and the very last elements of the whole sample play a special role, so
we should separately discuss the boundary conditions.

First of all, it is evident that the ${\frak r}_{+}$ can differ from the $%
{\frak r}_{-}$ at most by one unit. For ${\frak r}_{+}={\frak r}_{-}+1,$ the
sample starts and ends up with up-moves. We set for ${\frak r}_{+}={\frak %
r}_{-}+1$ the probability of finding the first move up equal to zero. For 
${\frak r}_{+}={\frak r}_{-}-1,$ the sample starts and ends up with down-moves.
For ${\frak r}_{+}={\frak r}_{-}-1$, the probability of finding
the first move down is set equal to zero either. For ${\frak r}_{+}=%
{\frak r}_{-},$ there are two samples, the one of which starts with an up-move,
and another one starts from a down-move. The probabilities of
finding the first elements in these two configurations are selected as would
the problem is formulated for a periodic chain: In the first case the
probability of the first element equals $1-q$, while in the second case the
probability of the first element equals $1-p$. Now, the problem is well
formulated. In what follows
we denote ${\frak r}={\frak r}_{+}={\frak r}_{-}$.

The boundary conditions are, clearly, not unique. We are interested,
however, by the limit of large $N$ where the effects of the boundary
conditions are not important. The above description is needed to formulate
unambiguously the problem.

Now, we are in a position to find the dispersion of the return $R_2$. Let us
sum up the probability $W({\frak r}_{+j},{\frak r}_{-j},N)$ over the numbers 
${\frak r}_{+j}$ and ${\frak r}_{-j}$ keeping, however, the values ${\frak r}%
_{+},$ ${\frak r}_{-},$ $n_{+},$ and $n_{-}$ fixed. It can be done with the
use of the method described in ref. \cite{Wilks}. We obtain 
\begin{equation}
W({\frak r}_{+j},{\frak r}_{-j},N)=C_N\left( 
\begin{array}{c}
n_{+}-1 \\ 
{\frak r}-1
\end{array}
\right) \left( 
\begin{array}{c}
n_{-}-1 \\ 
{\frak r}-1
\end{array}
\right) p^{n_{+}}q^{n_{-}}(\frac{1-q}p\frac{1-p}q)^{{\frak r}}.  \label{3.14}
\end{equation}
The asymptotic form at ${\frak r},$ $n_{+},$ and $n_{-}>>1$ of the $\ln W(%
{\frak r}_{+j},{\frak r}_{-j},N)$ is given by 
\begin{equation}
\ln W({\frak r}_{+j},{\frak r}_{-j},N)=-\frac{({\frak r}-<{\frak r}>)^2}{%
2\Sigma _{11}^2}-\frac{(n_{+}-<n_{+}>)^2}{2\Sigma _{22}^2}-\frac{({\frak r}-<%
{\frak r}>)(n_{+}-<n_{+}>)}{\Sigma _{12}^2}+...  \label{3.15}
\end{equation}
A simple calculation shows that $<{\frak r}>=<{\frak r}_{+}>=<{\frak r}_{-}>$%
, in agreement with eqs.(\ref{ar.p}) and (\ref{ar.m}), $<n_{+}>$ and $%
<n_{-}> $ are defined by eqs.(\ref{an.p})\ and (\ref{an.m}), and 
\begin{eqnarray}
\Sigma _{11}^2/N &=&\frac{pq(1-p)(1-q)}{(p+q)(2-p-q)},  \nonumber \\
\Sigma _{22}^2/N &=&\frac{pq(1-p)(1-q)}{((1-p)^2+(1-q)^2)(2-p-q)},  \nonumber
\\
\Sigma _{11}^2/N &=&\frac{pq}{(q-p)(2-p-q)}.  \label{3.16}
\end{eqnarray}
The dispersion of the value $R_1=2n_{+}-N$ is determined by the dispersion $%
\sigma _1^2$ of the value $n_{+}$, that equals 
\begin{equation}
\sigma _1^2=\frac{1/\Sigma _{11}^2}{1/(\Sigma _{11}^2\Sigma
_{22}^2)-1/\Sigma _{12}^4}.  \label{3.17}
\end{equation}
Notice that this value differs from the value of eq.(\ref{S1}) where the
two-step correlations are not taken into account. Respectively, 
\begin{equation}
\Delta R_1^2=4\sigma _1^2.  \label{3.18}
\end{equation}

The dispersion of the value $R_2=N-4r$ is determined by the dispersion $%
\sigma _2^2$ of the value ${\frak r}$: 
\begin{equation}
\sigma _2^2=\frac{1/\Sigma _{22}^2}{1/(\Sigma _{11}^2\Sigma
_{22}^2)-1/\Sigma _{12}^4}.  \label{3.17A}
\end{equation}
Respectively, 
\begin{equation}
\Delta R_2^2=16\sigma _2^2.  \label{3.18A}
\end{equation}

The problem is solved. It remains to compare the relative efficiency of
these two strategies. The results are placed in Table 6.

By comparing the values given in Table 6, one can
conclude that in the daily regime for all the indices the FT strategy is
more efficient, since it has better return with dispersion close to the
one of the alternative BH strategy. Here, however, transaction costs are
not taken into account.

{\tiny
\begin{table}
\caption{Comparison of the BH strategy ($\alpha =1$), FT strategy ($%
\alpha =0$), and the minimal risk strategy for the daily (D), weekly (W),
and monthly (M) quotes for the four major USA market indices. The parameter $%
\alpha $ is a mixing parameter for the combined strategy, $N$ is the number
of events (the number of the working days for the daily quotes, the number of
weeks and months for the weekly and monthly quotes). Return increases linearly
with $N$, while the dispersion increases as $\sqrt{N}$.}
\begin{center}
\begin{tabular}{|c|c|c|c|c|c|c|c|c|}
\hline\hline
Type & \multicolumn{2}{|c}{DJIA} & \multicolumn{2}{|c|}{S$\&$P} & 
\multicolumn{2}{|c|}{Nasdaq} & \multicolumn{2}{|c|}{NYSE} \\ 
\cline{2-3}\cline{4-5}\cline{6-7}\cline{8-9}
& $\alpha $ & Return $\pm $ dispersion & $\alpha $ & Return $\pm $ dispersion
& $\alpha $ & Return $\pm $ dispersion & $\alpha $ & Return $\pm $ dispersion
\\ \hline\hline
& $1$ & $0.0678N\pm 0.7880\sqrt{N}$ & $1$ & $0.0763N\pm 0.8334\sqrt{N}$ & $1$
& $0.1658N\pm 0.8909\sqrt{N}$ & $1$ & $0.0605N\pm 0.8456\sqrt{N}$ \\ 
D & $0$ & $0.0834N\pm 0.9932\sqrt{N}$ & $0$ & $0.1214N\pm 0.9887\sqrt{N}$ & $%
0$ & $0.2024N\pm 0.9642\sqrt{N}$ & $0$ & $0.1275N\pm 0.9894\sqrt{N}$ \\ 
& $0.6216$ & $0.0731N\pm 0.6114\sqrt{N}$ & $0.5830$ & $0.0952N\pm 0.6310%
\sqrt{N}$ & $0.5381$ & $0.1827N\pm 0.6429\sqrt{N}$ & $0.5767$ & $0.0889N\pm
0.6379\sqrt{N}$ \\ \hline\hline
& $1$ & $0.1245N\pm 0.7313\sqrt{N}$ & $1$ & $0.1452N\pm 0.7281\sqrt{N}$ & $1$
& $0.2637N\pm 0.7769\sqrt{N}$ & $1$ & $0.1159N\pm 0.7479\sqrt{N}$ \\ 
W & $0$ & $0.0529N\pm 0.9856\sqrt{N}$ & $0$ & $0.0595N\pm 0.9806\sqrt{N}$ & $%
0$ & $0.1375N\pm 0.9618\sqrt{N}$ & $0$ & $0.0628N\pm 0.9872\sqrt{N}$ \\ 
& $0.6397$ & $0.0984N\pm 0.5762\sqrt{N}$ & $0.6387$ & $0.1142N\pm 0.5718%
\sqrt{N}$ & $0.6060$ & $0.1772N\pm 0.5887\sqrt{N}$ & $0.6309$ & $0.0963N\pm
0.5859\sqrt{N}$ \\ \hline\hline
& $1$ & $0.1631N\pm 0.7312\sqrt{N}$ & $1$ & $0.1853N\pm 0.7953\sqrt{N}$ & $1$
& $0.3179N\pm 0.6130\sqrt{N}$ & $1$ & $0.1641N\pm 0.8057\sqrt{N}$ \\ 
M & $0$ & $0.0724N\pm 0.9757\sqrt{N}$ & $0$ & $0.1413N\pm 0.9667\sqrt{N}$ & $%
0$ & $0.1049N\pm 0.9038\sqrt{N}$ & $0$ & $0.1358N\pm 0.9726\sqrt{N}$ \\ 
& $0.6340$ & $0.1298N\pm 0.5711\sqrt{N}$ & $0.5915$ & $0.1676N\pm 0.6003%
\sqrt{N}$ & $0.6698$ & $0.2476N\pm 0.4841\sqrt{N}$ & $0.5893$ & $0.1525N\pm
0.6076\sqrt{N}$ \\ \hline\hline
\end{tabular}
\end{center}
\end{table}
}

One can expect that combining two strategies decreases the dispersion.
This is not always the case. Let us assume that funds invested to two
strategies are in the ratio $\alpha :\beta $, with $\alpha +\beta =1.$ The
return of the combined strategy equals
\begin{equation}
R_3=\alpha R_1+\beta R_2.  \label{3.19}
\end{equation}
In order to calculate dispersion, one should take into account the
correlation between the values $R_1$ and $R_2$. We obtain 
\begin{equation}
\Delta R_3^2=\frac{4\alpha ^2/\Sigma _{11}^2+16\beta ^2/\Sigma
_{22}^2+16\alpha \beta /\Sigma _{12}^2}{1/(\Sigma _{11}^2\Sigma
_{22}^2)-1/\Sigma _{12}^4}.  \label{3.20}
\end{equation}
The minimum of $\Delta R_3^2$ is achieved either at the boundaries $\alpha
=1 $ (BH strategy), $\alpha =0$ (FT strategy), or at that value of $%
\alpha $ where the first derivative of $\Delta R_3^2$ with respect to the $%
\alpha $ vanishes. In the last case, one should require (i) $0<\alpha <1$
and (ii) the second derivative is positive. In the case of a boundary
minimum, the combination of the two strategies does not result in a decrease
of the dispersion. In the second case, the dispersion can be decreased.

The combination of the BH and FT strategies appears to be constructive:
The conditions (i) and (ii) are satisfied. In Table 6, we give values of the
mixing parameter $\alpha ,$ the return $R_3$, and the dispersion of the
combined strategy with the lowest possible dispersion. The minimum of the 
dispersion is not an obligatory requirement. The value of $\alpha 
$ is selected as a compromise between the highest possible return and 
the minimum possible risk.
The above estimates show the limits within which these parameters (return
and dispersion) can be changed.

\section{Is large-scale dynamics determined by short-scale dynamics?}
\setcounter{equation}{0}
$\;$
\vspace{-0.5cm}

We can combine successive moves in pairs, triplets, and so on. The each
group of $k$ events ($k=1,2,...$) corresponds to a move up or down,
according as the total displacement of the index is the positive one or
the negative one. We wish to find parameters $W_k$, $P_k$, and $Q_k$ which describe
the probability of finding a group of the $k$ events in the state ''up'',
and, respectively, the conventional probabilities to find two such
successive groups in the states ''up'' and two successive groups in the
states ''down''.

The problem is motivated by the fact that participants of the market make
transactions with different frequencies: once per minute, once per hour, ...
, up to once per month and once per year. In the last case e.g. the change of the index
during one month can be treated as an elementary move. The problem of
finding parameters $W_k$, $P_k$, and $Q_k$ with $k>1$ in terms of the parameters 
$w$, $p$, and $q$ of the smallest time interval, has a unique
formal solution, as long as the model is clearly formulated. In the view of
the specific feature of the market, it is absolutely not apparent,
however, that the large-scale dynamics is determined by the short-scale
dynamics. The short-term traders take into account the behavior 
of the long-term traders, since the long-term traders are usually the 
institutional ones. From other side, the long-term investors take into account the
short-term behavior of the market.

Here, our purpose is to check to what extent the large-scale dynamics is determined by the 
short-scale dynamics. It can be done as follows: We
calculate parameters $W_k$, $P_k$, and $Q_k$ in terms of the parameters $w$, 
$p$, and $q$, determine parameters $W_k$, $P_k$, and $Q_k$ from the
historical quotes, and compare the two groups of the values.

\subsection{Weekly and monthly parameters in terms of daily parameters}

Let $k$ be the number of moves (days) in the group. The probability of
finding the group of $k$ moves ''up'' is given by 
\begin{equation}
W(p,q,k)=\sum_{{\frak r}_{+},{\frak r}_{-}}\sum_{n_{+}>n_{-}}W({\frak r}_{+},%
{\frak r}_{-},n_{+},n_{-})\gamma ({\frak r}_{+},{\frak r}_{-})  \label{4.1}
\end{equation}
where 
\begin{equation}
W({\frak r}_{+},{\frak r}_{-},n_{+},n_{-})=\left( 
\begin{array}{c}
n_{+}-1 \\ 
{\frak r}_{+}-1
\end{array}
\right) \left( 
\begin{array}{c}
n_{-}-1 \\ 
{\frak r}_{-}-1
\end{array}
\right) p^{n_{+}}q^{n_{-}}(\frac{1-q}p)^{{\frak r}_{+}}(\frac{1-p}q)^{{\frak %
r}_{-}}.  \label{4.2}
\end{equation}
The additional factor $\gamma ({\frak r}_{+},{\frak r}_{-})$, as compared to
eq.(\ref{W}), appears because of new boundary conditions. The first up event
comes with a probability $w$, the first event down comes with a probability $%
1-w$, ${\frak r}_{+}$ coincides with ${\frak r}_{-}$ or differs from ${\frak %
r}_{-}$ by one unit. The factor 
\begin{equation}
\gamma ({\frak r}_{+},{\frak r}_{-})=w\chi _{+}({\frak r}_{+},{\frak r}%
_{-})+(1-w)\chi _{-}({\frak r}_{+},{\frak r}_{-})  \label{4.3}
\end{equation}
where 
\begin{eqnarray}
\chi _{+}({\frak r}_{+},{\frak r}_{-}) &=&\left\{ 
\begin{array}{l}
\frac 1{1-q},\;if\;{\frak r}_{+}={\frak r}_{-}+1\;or\;{\frak r}_{+}={\frak r}%
_{-} \\ 
0,\;\;\;\;if\;{\frak r}_{+}={\frak r}_{-}-1
\end{array}
\right.  ,\label{4.4} \\
\chi _{-}({\frak r}_{+},{\frak r}_{-}) &=&\left\{ 
\begin{array}{l}
0,\;\;\;\;if\;{\frak r}_{+}={\frak r}_{-}+1 \\ 
\frac 1{1-p},\;if\;{\frak r}_{+}={\frak r}_{-}-1\;or\;{\frak r}_{+}={\frak r}%
_{-}
\end{array}
\right.  .\label{4.5}
\end{eqnarray}
The first term in eq.(4.3) corresponds to the case when in the group of $k$
events the first move is an up-move, while the second term corresponds to
the case of a down-move. In eq.(4.1), respectively, $n_{+}$ and $n_{-}$
are the numbers of the moves up and down in the group. Apparently, $%
n_{+}+n_{-}=k.$ The values ${\frak r}_{+}$ and ${\frak r}_{-}$ give the
number of the series up and down in the considered group.

For an even $k,$ a situation is possible when total displacement is equal to
zero, $n_{+}=n_{-}$. In such a case, it is necessary to specify separately the
meaning of the up and down moves for groups of the $k$ events. We assign the
probability of $1/2$ for interpreting neutral move as a move up and the
probability of $1/2$ for interpreting neutral move as a move down. For
even $k,$ the summation in eq.(4.1) extends to $n_{+}=n_{-}$ with the weight
of $1/2$. The normalization factor in eq.(4.1) should be equal to unity
according to the construction.

The probability that the second group of $k$ events has the value ''up'',
provided that the first group has the value ''up'' also, can be found from
equation 
\begin{eqnarray}
P(p,q,k) &=&\frac{W^{++}(p,q,k)}{W(p,q,k)}\sum_{{\frak r}_{+},{\frak r}%
_{-}}\sum_{n_{+}>n_{-}}W({\frak r}_{+},{\frak r}_{-},n_{+},n_{-})\chi _{+}(%
{\frak r}_{+},{\frak r}_{-})  \nonumber \\
&&\ +\frac{W^{+-}(p,q,k)}{W(p,q,k)}\sum_{{\frak r}_{+},{\frak r}%
_{-}}\sum_{n_{+}>n_{-}}W({\frak r}_{+},{\frak r}_{-},n_{+},n_{-})\chi _{-}(%
{\frak r}_{+},{\frak r}_{-})  \label{4.6}
\end{eqnarray}
where the ratios $W^{++}(p,q,k)/W(p,q,k)$ and $W^{+-}(p,q,k)/W(p,q,k)$ are
the probabilities that the last events in the up-groups are, respectively,
moves up and down. The values $W^{++}(p,q,k)$ and $W^{+-}(p,q,k)$ are
calculated as follows: 
\begin{eqnarray}
W^{++}(p,q,k) &=&\sum_{{\frak r}_{+},{\frak r}_{-}}\sum_{n_{+}>n_{-}}W(%
{\frak r}_{+},{\frak r}_{-},n_{+},n_{-})\eta _{+}({\frak r}_{+},{\frak r}%
_{-}),  \label{4.7} \\
W^{+-}(p,q,k) &=&\sum_{{\frak r}_{+},{\frak r}_{-}}\sum_{n_{+}>n_{-}}W(%
{\frak r}_{+},{\frak r}_{-},n_{+},n_{-})\eta _{-}({\frak r}_{+},{\frak r}%
_{-})  \label{4.8}
\end{eqnarray}
where 
\begin{equation}
\eta _{+}({\frak r}_{+},{\frak r}_{-})=\left\{ 
\begin{array}{l}
\frac w{1-q},\;if\;{\frak r}_{+}={\frak r}_{-}+1 \\ 
\frac{1-w}{1-p},\;if\;{\frak r}_{+}={\frak r}_{-} \\ 
0,\;\;\;\;if\;{\frak r}_{+}={\frak r}_{-}-1
\end{array}
\right. ,  \label{4.9}
\end{equation}
\begin{equation}
\eta _{-}({\frak r}_{+},{\frak r}_{-})=\left\{ 
\begin{array}{l}
0,\;\;\;\;if\;{\frak r}_{+}={\frak r}_{-}+1 \\ 
\frac w{1-q},\;if\;{\frak r}_{+}={\frak r}_{-} \\ 
\frac{1-w}{1-p},\ if\;{\frak r}_{+}={\frak r}_{-}-1
\end{array}
\right. .  \label{4.10}
\end{equation}
Eq.(4.7) can be explained in the following way: The last move in the group
of $k$ moves is a move up provided that ${\frak r}_{+}={\frak r}_{-}+1.$
Else, it is a move up also at ${\frak r}_{+}={\frak r}_{-}$ provided that
the first move in the group is a move down. In the first case, the first
move is a move up, and so the probability $W({\frak r}_{+},{\frak r}%
_{-},n_{+},n_{-})$ should be divided by $1-q$ and multiplied by $w$. In the
second case, respectively, the probability should be divided by $1-p$ and
multiplied by $1-w$.

The value $Q(p,q,k)$ can be found in the similar way:

\begin{eqnarray}
Q(p,q,k) &=&\frac{W^{-+}(p,q,k)}{1-W(p,q,k)}\sum_{{\frak r}_{+},{\frak r}%
_{-}}\sum_{n_{+}<n_{-}}W({\frak r}_{+},{\frak r}_{-},n_{+},n_{-})\chi _{+}(%
{\frak r}_{+},{\frak r}_{-})  \nonumber \\
&&\ +\frac{W^{--}(p,q,k)}{1-W(p,q,k)}\sum_{{\frak r}_{+},{\frak r}%
_{-}}\sum_{n_{+}<n_{-}}W({\frak r}_{+},{\frak r}_{-},n_{+},n_{-})\chi _{-}(%
{\frak r}_{+},{\frak r}_{-})  \label{4.11}
\end{eqnarray}
where 
\begin{eqnarray}
W^{-+}(p,q,k) &=&\sum_{{\frak r}_{+},{\frak r}_{-}}\sum_{n_{+}<n_{-}}W(%
{\frak r}_{+},{\frak r}_{-},n_{+},n_{-})\eta _{+}({\frak r}_{+},{\frak r}%
_{-}),  \label{4.12} \\
W^{--}(p,q,k) &=&\sum_{{\frak r}_{+},{\frak r}_{-}}\sum_{n_{+}<n_{-}}W(%
{\frak r}_{+},{\frak r}_{-},n_{+},n_{-})\eta _{-}({\frak r}_{+},{\frak r}%
_{-})  .\label{4.13}
\end{eqnarray}

It is evident that 
\begin{eqnarray}
W^{++}(p,q,k)+W^{+-}(p,q,k) &=&W(p,q,k),  \label{4.14} \\
W^{-+}(p,q,k)+W^{--}(p,q,k) &=&1-W(p,q,k).  \label{4.15}
\end{eqnarray}
One can show that the homogeneous condition (\ref{homo}) is satisfied for
the values $W$, $P$, and $Q$.

Since the algorithm for construction of the quantities $W$, $P$, and $Q$ is
simple, the numerical solution can be found quite easily.

In Table 1, we place the calculated values $W$, $P$, and $Q$ for $k=5$
(weekly quotes) and $21$ (monthly quotes), the values $w$, $p$, and $q$ are
determined from fitting the daily quotes. We show also the empirical values
of $W$, $P$, and $Q$ determined from the historical weekly and monthly quotes.
In Table 2, the values of the $\chi ^2$ are shown for the empirical weekly
and monthly quotes for three models of the random walking, discussed in the
previous Sect. These quotes are analyzed in the same way as the daily quotes.

The comparison of the calculated and empirical values $W$, $P$, and $Q$
shows quite good agreement for the weekly quotes. A noticeable deviation
exists only for Nasdaq in the value $Q_5$. When the index falls down, a
stronger than expected correlation exists: The probability of the
continuation of the down-moves exceeds the calculated value. The system easily 
evolves from "bad state to worse state" (Parkinson Law).

In all other cases, the agreement of the weekly parameters $W$, $P$, and $Q$
with the calculated values is satisfactory. One can conclude that at the
daily-weekly scale, the short-scale dynamics (daily behavior of the indices)
determines the large-scale dynamics (the weekly data).

Let us compare now the calculated monthly parameters $W$, $P$, and $Q$ with
the empirical values. We see that the values $Q_{21}$'s are essentially
underestimated for all indices. We observed such an effect already for
the weekly quotes of Nasdaq. In the monthly quotes, the effect is much more pronounced.
At the same time, the calculated values $W$ and $P$ coincide within one-two
standard deviations with the empirical values.

We conclude that the large-scale dynamics (monthly behavior) is determined
by the short-scale dynamics (daily behavior) only partially. 
It is quite different from what we have in physics. On the monthly
scale, there are substantial deviations from the random walking if the
memory on the past is, as we assumed, restricted by the previous day only. 
It is also interesting that the values $W$ and $P$ calculated
within the framework of such a simple model are in reasonable agreement 
with the empirical values.

\subsection{Renorm group in random walking problem}

There is a renorm group relation between parameters $W$, $P$, and $Q$: 
\begin{eqnarray}
P(P(p,q,l),Q(p,q,l),k) &=&P(p,q,kl),  \label{5.1} \\
Q(P(p,q,l),Q(p,q,l),k) &=&Q(p,q,kl).  \label{5.2}
\end{eqnarray}
These relations can be used to establish the form of the functions $P$ and $%
Q $ with the help of the boundary conditions 
\begin{eqnarray}
P(p,q,1) &=&p,  \label{5.3} \\
Q(p,q,1) &=&q.  \label{5.4}
\end{eqnarray}
The algorithm described in the previous Sect. can be used to construct
functions $P$ and $Q$ for positive integer $k$. In the renorm group
equations, we can set $k=1/l$ and find functions $P$ and $Q$ for $k=1/l$. We
set afterwards $k=1/m$ where $m$ is an arbitrary integer number and obtain
the functions $P$ and $Q$ for all rational values of the arguments.

\section{Combining Series Analysis with Candlesticks Charting Techniques}
\setcounter{equation}{0}
$\;$
\vspace{-0.5cm}

The BH and FT strategies have comparable
dispersions for the analyzed indices. These strategies are, however, very
different. This is why the returns do not correlate strongly. It means that
combining these two strategies is effective to reduce the dispersion.

In this Sect., we attempt to develop a computer-based strategy which 
is distinct from the previous ones. This strategy uses new technical 
tools. Due to this reason, we expect that its return does not correlate 
strongly with the BH and FT returns, so that by combining these strategies 
the dispersion can further be reduced.

The candlesticks charting techniques is believed to be useful for
understanding the major market moves and its turning points \cite{Nison}. It is
the most useful when it is supplemented with general analysis of the
market. The series analysis which we discussed in the
previous Sects. provides useful hints in this respect. Long series e.g.
result into overbought or oversold states, in which the candlesticks
patterns work differently. We perform a backtest of the candlesticks
patterns separately for the up- and down-series of each length. One can
expect that combining the series analysis with the candlesticks techniques
can improve the record of the single methods.

{\tiny
\begin{table}
\caption{The year-by-year performance of
the BH strategy, FT strategy, and the minimum risk strategy (MR) for the
daily DJIA index. We show also
the performance of the computer-based strategy (CB) described in this Sect. In the
last two columns, the number of transactions per year (TR) and the win-to-loss
ratio (W/L) are given for the CB strategy.}
\begin{center}
\begin{tabular}{||c|c|c|c|c|c|c||c|c|c|c|c|c|c||}
\hline\hline
Year & BH & FT & MR & CB & TR & W/L     & Year & BH  & FT & MR & CB & TR & W/L  \\ \hline\hline
1936 & 34 & -13 & 15 & -4 & 19 & 0.6522 & 1968 & 9   & 35 & 19 & 8 & 95 & 1.1839 \\
1937 & -1 & 12 & 4 & 3 & 23 & 1.3000    & 1969 & -20 & 37 & 2 & 5 & 72 & 1.1493 \\
1938 & 4  & -10 & -1 & -1 & 57 & 0.9655 & 1970 & 1   & 42 & 17 & 14 & 61 & 1.5957 \\
1939 & 2  & -13 & -4 & 4 & 26 & 1.3636  & 1971 & 12  & 46 & 25 & 22 & 67 & 1.9778 \\
1940 & 15 & 9 & 12 & -2 & 23 & 0.8400   & 1972 & -6  & 36 & 10 & 8 & 83 & 1.2133 \\
1941 & -18 & 44 & 6 & 4 & 19 & 1.5333   & 1973 & -29 & 53 & 3 & 10 & 114 & 1.1923 \\ 
1942 & 30 & 37 & 32 & 2 & 10 & 1.5000   & 1974 & -34 & 50 & 0 & -1 & 73 & 0.9730 \\
1943 & 33 & 53 & 41 & -4 & 35 & 0.7949  & 1975 & 17  & 33 & 23 & 0 & 12 & 1.0000 \\ 
1944 & 48 & 25 & 38 & 1 & 11 & 1.2000   & 1976 & 8   & 0  & 4 & 8 & 82 & 1.2162 \\
1945 & 40 & 58 & 47 & -3 & 31 & 0.8235  & 1977 & -7  & 19 & 3 & -2 & 85 & 0.9540 \\
1946 & -4 & 58 & 20 & 18 & 87 & 1.5217  & 1978 & 2   & 25 & 11 & 20 & 114 & 1.4255  \\
1947 & 15 & 20 & 17 & 20 & 73 & 1.7547  & 1979 & 16  & -10 & 5 & -8 & 109 & 0.8632  \\
1948 & 18 & 14 & 16 & 16 & 75 & 1.5424  & 1980 & 27  & -2 & 15 & 3 & 91 & 1.0682  \\
1949 & 18 & 17 & 17 & 27 & 90 & 1.8571  & 1981 & -16 & 8 & -6 & 0 & 80 & 1.0000  \\
1950 & 53 & 27 & 42 & 27 & 84 & 1.9474  & 1982 & -16 & 4 & -8 & -1 & 67 & 0.9706  \\
1951 & 11 & 26 & 17 & 19 & 89 & 1.5429  & 1983 & 19  & 4 & 13 & 5 & 108 & 1.0971  \\
1952 & 23 & 44 & 31 & 15 & 84 & 1.4348  & 1984 & -26 & 12 & -10 & 2 & 102 & 1.0400  \\
1953 & 7 & 38 & 19 & 24 & 70 & 2.0435   & 1985 & 19  & -7 & 8 & 14 & 93 & 1.3544  \\
1954 & 67 & 47 & 59 & 28 & 103 & 1.7467 & 1986 & 22  & -10 & 9 & 5 & 72 & 1.1493  \\
1955 & 58 & 45 & 52 & 21 & 75 & 1.7778  & 1987 & 34  & 20 & 28 & 6 & 58 & 1.2308  \\
1956 & 2 & 40 & 17 & 16 & 91 & 1.4267   & 1988 & 15  & -22 & 0 & 7 & 73 & 1.2121  \\
1957 & -3 & 19 & 5 & 14 & 93 & 1.3544   & 1989 & 31  & 7 & 21 & 11 & 64 & 1.4151  \\
1958 & 42 & 35 & 39 & 28 & 74 & 2.2174  & 1990 & 15  & -6 & 6 & 12 & 95 & 1.2892  \\
1959 & 38 & 34 & 36 & 20 & 107 & 1.4598 & 1991 & -3  & 30 & 10 & 9 & 76 & 1.2687  \\
1960 & -15 & 53 & 12 & 13 & 102 & 1.2921& 1992 & 3   & -29 & -9 & -14 & 70 & 0.6667  \\
1961 & 23 & 17 & 20 & 11 & 74 & 1.3492  & 1993 & 39 & -26 & 13 & 18 & 68 & 1.7200  \\
1962 & -12 & 31 & 5 & -5 & 89 & 0.8936  & 1994 & 5 & 9 & 6 & 0 & 0 & 0.0000  \\
1963 & 25 & 2 & 15 & 5 & 98 & 1.1075    & 1995 & 33 & 7 & 22 & 11 & 67 & 1.3929  \\
1964 & 39 & 36 & 37 & 15 & 71 & 1.5357  & 1996 & 39 & 11 & 27 & 9 & 60 & 1.3529  \\
1965 & 30 & 26 & 28 & 13 & 64 & 1.5098  & 1997 & 23 & 0 & 13 & 0 & 4 & 1.0000  \\
1966 & -27 & 59 & 7 & 9 & 55 & 1.3913   & 1998 & 21 & 23 & 21 & 3 & 11 & 1.7500  \\
1967 & 13 & 46 & 26 & 19 & 69 & 1.7600  & 1999 & -2 & -17 & -8 & -1 & 68 & 0.9710 \\ 
\hline\hline
\end{tabular}
\end{center}
\end{table}
}

\subsection{Use and misuse of computer backtesting}

Backtesting means the following: Suppose we have a well fixed strategy, i.e.
a fixed set of prescriptions when to buy and sell an index. If we
have historical quotes, the performance of the strategy can be tested. Such a
procedure is called backtesting. It gives an idea on the possible future
performance.

We make distinction between causal backtesting and non-causal backtesting. 
To explain the difference, consider an example: Let we work with a moving average which
has just one free parameter, the number of days to be averaged. This
parameter can be treated as a fitting parameter. It can be determined by
requiring that the return during the backtested period be the greatest one.

The return obtained this way is, however, {\it not} the one to be expected from the
future performance. Since the fitted
parameter depends on the whole backtested set of the historical quotes, the
return e.g. in the year 1975 depends e.g. on the year 1995, in the apparent
contradiction with the causality. The non-causal backtesting
gives an idea on the reasonable values of the fitting parameters, but the return
is always overestimated. The example described above 
illustrates the possible misuse of the computer backtesting.

{\tiny
\begin{table}
\caption{The year-by-year performance of
the BH strategy, FT strategy, and the minimum risk strategy (MR) for the
daily S$\&$P index. We show also
the performance of the computer-based strategy (CB) described in this Sect. In the
last two columns, the number of transactions per year (TR) and the win-to-loss
ratio (W/L) are given for the CB strategy.}
\begin{center}
\begin{tabular}{||c|c|c|c|c|c|c||c|c|c|c|c|c|c||}
\hline\hline
Year & BH & FT & MR & CB & TR & W/L     & Year & BH & FT & MR & CB & TR & W/L \\\hline\hline
1955 & -2 & -17 & -8 & -1 & 68 & 0.9710 & 1978 & 9 & 43 & 22 & 28 & 145 & 1.4786 \\ 
1956 & -15 & 23 & 0 & -10 & 62 & 0.7222 & 1979 & 32 & 25 & 29 & 16 & 128 & 1.2857 \\
1957 & -9 & 44 & 12 & 1 & 79 & 1.0256   & 1980 & 29 & 38 & 32 & 8 & 102 & 1.1702 \\
1958 & 63 & 47 & 56 & 28 & 68 & 2.4000  & 1981 & -11 & 28 & 4 & 5 & 162 & 1.0637 \\  
1959 & 23 & 16 & 20 & 10 & 52 & 1.4762  & 1982 & -25 & 8 & -11 & -10 & 139 & 0.8658 \\  
1960 & -3 & 49 & 17 & 7 & 47 & 1.3500   & 1983 & 23 & -12 & 9 & 7 & 102 & 1.1474 \\   
1961 & 49 & 27 & 40 & 18 & 50 & 2.1250  & 1984 & -27 & 10 & -12 & -3 & 41 & 0.8636 \\  
1962 & -13 & 39 & 7 & 7 & 55 & 1.2917   & 1985 & 22 & 15 & 19 & -2 & 27 & 0.8621 \\ 
1963 & 39 & 22 & 32 & 0 & 0 & 0.0000    & 1986 & 29 & -4 & 15 & 9 & 33 & 1.7500 \\   
1964 & 44 & 42 & 43 & 0 & 0 & 0.0000    & 1987 & 33 & 18 & 27 & -3 & 31 & 0.8235 \\  
1965 & 36 & 41 & 38 & 0 & 0 & 0.0000    & 1988 & 25 & -14 & 9 & 1 & 19 & 1.1111 \\  
1966 & -10 & 69 & 21 & 0 & 0 & 0.0000   & 1989 & 48 & 13 & 34 & 4 & 30 & 1.3077 \\
1967 & 35 & 44 & 38 & 5 & 13 & 2.2500   & 1990 & 17 & 4 & 11 & -1 & 37 & 0.9474 \\ 
1968 & 9 & 43 & 22 & 6 & 20 & 1.8571    & 1991 & -5 & 6 & 0 & 0 & 38 & 1.0000 \\ 
1969 & -4 & 63 & 22 & 1 & 11 & 1.2000   & 1992 & 7 & -9 & 0 & 4 & 28 & 1.3333 \\  
1970 & 2 & 45 & 19 & -1 & 10 & 0.8182   & 1993 & 7 & 24 & 13 & 2 & 32 & 1.1333 \\   
1971 & 23 & 56 & 36 & 5 & 13 & 2.2500   & 1994 & 14 & -13 & 3 & 2 & 28 & 1.1538 \\ 
1972 & 26 & 58 & 38 & 4 & 55 & 1.1569   & 1995 & 60 & 21 & 44 & 18 & 58 & 1.9000 \\  
1973 & -26 & 47 & 3 & 4 & 51 & 1.1702   & 1996 & 22 & 27 & 24 & 13 & 60 & 1.5532 \\  
1974 & -41 & 48 & -5 & 30 & 118 & 1.6818  &  1997 & 28 & 8 & 20 & 6 & 64 & 1.2069 \\  
1975 & 18 & 34 & 24 & 6 & 58 & 1.2308   & 1998 & 28 & 9 & 20 & -6 & 81 & 0.8621 \\ 
1976 & 19 & 14 & 17 & 5 & 129 & 1.0806  & 1999 & 6 & 1 & 4 & 11 & 81 & 1.3143 \\ 
1977 & -14 & 29 & 3 & 10 & 92 & 1.2439  &      &   &   &   &    &    &        \\ 
\hline\hline
\end{tabular}
\end{center}
\end{table}
}

Causal backtesting means the following: In the year 1975, we calculate from the
historical quotes an optimal value of the days for a moving average. Using this
optimal value, we calculate then the return in the year 1976. After that, we
calculate an optimal value of the days by adding the year 1976 to the set of the historical
quotes, which we used earlier to fix the fitting parameter. This new optimal
value is used to calculate the return in the year 1977. The procedure is
repeated every year. The total return is a sum of the returns calculated for
all the years. Such a calculation is essentially equivalent to a real-time
testing. The causality, clearly, is not violated, since the calculated
return e.g. in the year 1977 depends on the historical quotes from the
previous years only. The return from the causal
backtesting gives more realistic idea on the future performance. It is always lower than
the return form the non-causal backtesting. In what
follows, we speak on the causal backtesting only.

\subsection{Analytic images of candlesticks charting patterns for computers}

The characteristic candlesticks patterns are described in the literature
in an intuitive fashion \cite
{Nison}. In order to implement the candlesticks charting techniques 
into a computer code, one needs to formulate in a language clear for 
every programmer how to distinguish the special informative patterns from the common ones. 
We include into our code about 30 analytic images which correspond to about 20
typical candlesticks patterns. Since clear prescriptions on how to construct the analytic
images do not exist, several analytic images are
usually probed. The code by analyzing the past performance
decides which analytic images work better.

{\tiny
\begin{table}
\caption{The year-by-year performance of
the BH strategy, FT strategy, and the minimum risk strategy (MR) for the
daily Nasdaq index. We show also
the performance of the computer-based strategy (CB) described in this Sect. In the
last two columns, the number of transactions per year (TR) and the win-to-loss
ratio (W/L) are given for the CB strategy.}
\begin{center}
\begin{tabular}{||c|c|c|c|c|c|c||}
\hline\hline
Year & BH & FT & MR & CB & NT & W/L \\ \hline\hline
1990 & 6 & 1 & 4 & 11 & 81 & 1.3143 \\ 
1991 & 45 & 58 & 50 & 50 & 183 & 1.7519 \\ 
1992 & 12 & 37 & 22 & 31 & 177 & 1.4247 \\ 
1993 & 57 & 50 & 54 & 43 & 146 & 1.8350 \\ 
1994 & 13 & 39 & 23 & 26 & 164 & 1.3768 \\ 
1995 & 44 & 37 & 41 & 40 & 151 & 1.7207 \\ 
1996 & 48 & 37 & 43 & 40 & 167 & 1.6299 \\ 
1997 & 37 & 46 & 40 & 57 & 178 & 1.9421 \\ 
1998 & 38 & 19 & 30 & 11 & 174 & 1.1350 \\ 
1999 & 40 & 7 & 26 & 28 & 137 & 1.5138 \\ \hline\hline
\end{tabular}
\end{center}
\end{table}
}

{\bf Example 1}: The one-line "hammer" pattern is expected to be bullish
in the downtrend. This pattern
does work in down-series. Several analytic images can be proposed to
distinguish the hammer pattern:

\begin{eqnarray}
\xi &=&\frac{H-L}{10\min (H-O,H-C)+\max (H-O,H-C)}, \\
\xi &=&\frac{H-L}{10(H-C)+\left| C-O\right| },\;etc
\end{eqnarray}
where $H$, $L$, $O$, $C$ are the high, low, open, and close of the index.
For the hammer pattern, the value of $\xi $ is large positive. The code uses
this signature to recognize the hammer pattern between the others. How large $%
\xi $ should be is decided by computer from the analysis of the historical
performance of the considered image. The critical value of $\xi $ depends on
the length of the down-series in which the bullish pattern appears. 

{\bf Example 2}: Two-line candlesticks pattern "bullish engulfing
lines". The following analytic image can be proposed 
\begin{equation}
\xi =\frac{H-L}{<H-L>}\frac{\left| C-O\right| }{H-L-\left| C-O\right| }\frac{%
\left| C-O\right| -\left| C^{\prime }-O^{\prime }\right| }{<H-L>},
\end{equation}
if $(C^{\prime }-O^{\prime })(C-O)<0$ and $\xi =0$ otherwise. Here, $%
C^{\prime }$ and $O^{\prime }$ are the close and open of the analyzed index
from the previous day. The values of $\xi $ should be large when the bullish
engulfing lines pattern is formed. How large the value of $\xi $ should be
is decided by the computer from analysis of the past pattern performance.
The critical value of $\xi $ are different for the down-series of the
different lengths, in which the bullish pattern appears.

{\tiny
\begin{table}
\caption{The year-by-year performance of
the BH strategy, FT strategy, and the minimum risk strategy (MR) for the
daily NYSE index. We show also
the performance of the computer-based strategy (CB) described in this Sect. In the
last two columns, the number of transactions per year (TR) and the win-to-loss
ratio (W/L) are given for the CB strategy.}
\begin{center}
\begin{tabular}{||c|c|c|c|c|c|c||}
\hline\hline
Year & BH & FT & MR & CB & NT & W/L \\ \hline\hline
1971 & 40 & 7 & 26 & 28 & 137 & 1.5138 \\ 
1972 & 24 & 60 & 38 & 50 & 195 & 1.6897 \\ 
1973 & -28 & 53 & 4 & 50 & 189 & 1.7194 \\ 
1974 & -41 & 52 & -3 & 35 & 190 & 1.4516 \\ 
1975 & 10 & 34 & 19 & 25 & 215 & 1.2632 \\ 
1976 & 18 & 22 & 19 & 24 & 234 & 1.2286 \\ 
1977 & -13 & 23 & 1 & 16 & 196 & 1.1778 \\ 
1978 & 12 & 57 & 30 & 40 & 180 & 1.5714 \\ 
1979 & 42 & 34 & 38 & 33 & 216 & 1.3607 \\ 
1980 & 30 & 42 & 34 & 34 & 196 & 1.4198 \\ 
1981 & -2 & 28 & 10 & 28 & 202 & 1.3218 \\ 
1982 & -28 & 16 & -10 & 12 & 186 & 1.1379 \\ 
1983 & 18 & -12 & 6 & -15 & 160 & 0.8286 \\ 
1984 & -19 & 18 & -4 & 15 & 109 & 1.3191 \\ 
1985 & 26 & 25 & 25 & 6 & 127 & 1.0992 \\ 
1986 & 35 & 0 & 21 & -4 & 194 & 0.9596 \\ 
1987 & 32 & 20 & 27 & -3 & 91 & 0.9362 \\ 
1988 & 28 & -2 & 16 & 5 & 135 & 1.0769 \\ 
1989 & 46 & 13 & 32 & -7 & 81 & 0.8409 \\ 
1990 & 23 & 20 & 21 & 14 & 131 & 1.2393 \\ 
1991 & 4 & 12 & 7 & 8 & 147 & 1.1151 \\ 
1992 & 9 & 9 & 9 & -1 & 74 & 0.9733 \\ 
1993 & 14 & 24 & 18 & 15 & 115 & 1.3000 \\ 
1994 & 10 & -1 & 5 & 2 & 115 & 1.0354 \\ 
1995 & 61 & 27 & 47 & 5 & 109 & 1.0962 \\ 
1996 & 20 & 31 & 24 & 22 & 143 & 1.3636 \\ 
1997 & 26 & 0 & 15 & 15 & 166 & 1.1987 \\ 
1998 & 28 & 19 & 24 & 11 & 116 & 1.2095 \\ 
1999 & -14 & 13 & -3 & 9 & 128 & 1.1513 \\ \hline\hline
\end{tabular}
\end{center}
\end{table}
}

The similar formulae can also be written for other patterns. Notice again
that there is no a unique prescription how to construct the analytic
images. The similar situation exists in the mathematical statistics where
the so-called ''statistics'', which play an important role in the decision
theory, are not uniquely defined. Nevertheless, they are widely used in the
data analysis (see e.g. \cite{Wilks,Portenko,Krivoruchenko} and references therein).

The analytic images work in the code as follows: We consider the first
5 to 10 years of the historical quotes to fix confidence intervals for the analytic 
images. We list and
keep information on the all up- and down-series of a fixed length. For each
series and each pattern, a distribution function of $\xi $ is constructed.
We compare the number of points in the confidence interval $\xi \in (-\infty
,a)$ which are accompanied by continuation and reversal of the
series. The value of $a$ is varied. We are looking for those values of $a$
which give the best chance of the continuation or the reversal of the
series. We pay attention to an image if and only if the confidence interval
provides a chance better than $3:1$ (win-to-loss ratio). There are many
confidence intervals providing such chances and containing many
data points. The win-to-loss ratio at this stage refers, however, 
to the non-causal backtesting. It should be overestimated, as we discussed above.
Indeed, the causal backtesting of the years, following the first ones,
shows that the actual win-to-loss (W/L) ratio is about 4:3, as shown
in Tables 7 - 10. The confidence intervals $\xi \in (b,+\infty )$ are
constructed in the same way and the common results are the similar.

The backtest results reported in Tables 7, 8, 9, and 10 for the computer-based 
strategy are mixed at present. There are extended time periods in the S$\&$P and NYSE
indices where the computer-based strategy worked well. Nasdaq gives also
good record. There are periods at DJIA, S$\&$P, and NYSE indices, however,
where the records are quite poor. The available code will be improved.

\section{Conclusion}
\setcounter{equation}{0}
$\;$
\vspace{-0.5cm}

In this paper, a statistical analysis of the major USA market indices DJIA, S$\&$P, 
Nasdaq, and NYSE is made. We verified that the market behavior is very close to the 
stochastic one and that the market dynamics is quite similar to the random walking. 

The Markov chains approach was applied to study two-step correlations in the 
daily, weekly, and monthly historical quotes. The correlation parameters are determined
empirically. The correlations are found to 
be statistically significant for the daily quotes, whereas the weekly and monthly 
quotes do not display clear effect.

There are deviations from the geometrical distribution of the number of series 
of the fixed length. The empirical probabilities are collected in Tables 3, 4, and 5.
There are exceptional deviations: Nasdaq during the years 1985 - 2000 did not 
fall off more than 6 weeks in raw in the weekly quotes and more than 4 month in 
raw in the monthly quotes.

We calculated the return and the dispersion of the buy-and-hold and follow-trend strategies. 
These strategies are combined to decrease the dispersion. The optimal ratio between
investments into these two strategies, providing a minimal risk, is found analytically.

The weekly and monthly correlation parameters are calculated in terms of the 
daily parameters. The results show reasonable agreement with the empirical data for
the weekly quotes. There are, however, evident deviations in the monthly quotes for
the down-series. The empirical probabilities for continuation of the down-trend
are noticeably greater than the calculated probabilities (Parkinson Law?). 
It can be interpreted to mean that the large-scale market dynamics is not fully 
determined by the short-scale dynamics.

It was shown that the correlation parameters at different scales obey the renorm
group equations. 

This work reported also the intermediate results of the constructing a computer-based 
strategy that combines the series analysis with the candlesticks charting techniques. 

\vspace{1.5 cm}
\begin{center}
{\large {\bf Acknowledgments}}
\end{center}

The results of the statistical analysis of the historical quotes for the four major
USA market indices DJIA, S$\&$P, Nasdaq, and NYSE are reported from 
the permission of Commodity Systems, Inc. (CSI), 
the copyright owner of these quotes (web-site http://www.csidata.com). The author
wishes to thank also the Dow Jones Indexes, the owner of the Dow Jones Industrial 
Average, for providing him with the DJIA historical quotes and the permission to use these 
quotes for the present publication. The author is grateful to European Physical Society
for a grant which made possible his participation at the Conference "Application of Physics 
in Financial Analysis" (5 - 7 December 2001, London).
The author is indebted to the Institute for Theoretical Physics 
of University of Tuebingen for kind hospitality and providing an opportunity to bring
this paper into the final form.

\newpage
%\begin{references}


\begin{thebibliography}{999}

\bibitem{Chandrasekhar}  S. Chandrasekhar, {\it Stochastic Problems in
Physics and Astronomy}, International Lit., Moscow, 1947.

\bibitem{Wilks}  C. Wilks, {\it Mathematical Statistics}, Nauka, Moscow,
1967.

\bibitem{Portenko}  V. S. Korolyuk, N. I. Portenko, A. V. Skorokhod, A. F.
Turbin, {\it Handbook on Probability Theory and Mathematical Statistics},
Nauka, Moscow, 1985.

\bibitem{Colby}  R. W. Colby and Th. A. Meyers, {\it The Encyclopedia of
Technical Market Indicators}, McGraw-Hill, New York e. a., 1988.

\bibitem{Nison}  S. Nison, {\it Beyond Candlesticks, New Japanese Charting
Techniques}, John Wiley $\&$ Sons, Inc., New York e. a., 1994.

\bibitem{Krivoruchenko}  M. I. Krivoruchenko, {\it Statistical Analysis of
the Angular Distribution of Neutrino Events Observed in Kamiokande II and
IMB Detectors from Supernova SN 1987 A}, Z. Phys. {\bf C44}, 633 (1989).

%\end{references}
\end{thebibliography}
\end{document}